\documentclass[pra, showpacs, twocolumn,amsmath,amssymb]{revtex4}

\usepackage{mathrsfs}
\usepackage{graphicx}
\usepackage{dcolumn}
\usepackage{bm}

\begin{document}


\title{Upper Bound on our Knowledge about Noncommuting Observables for a Qubit System}

\author{Yuji Kurotani$^1$}
\author{Takahiro Sagawa$^1$}
\author{Masahito Ueda$^{1,2}$}
\affiliation{$^1$ Department of Physics, Tokyo Institute of Technology, Tokyo 152-8551, Japan\\
$^2$ ERATO Macroscopic Quantum Control Project, JST, Tokyo 113-8656, Japan}

\date{\today}


\begin{abstract}
A trade-off relation on our knowledge about two noncommuting observables of a qubit system in simultaneous measurement is formulated. The obtained inequality offers a quantitative information-theoretic representation of Bohr's principle of complementarity, and can be interpreted as a trade-off relation on the asymptotic accuracy of the maximum-likelihood estimation of the probability distributions of observables. 
\end{abstract}

\maketitle

Quantum mechanics features two distinct kinds of uncertainty.
One is the quantum fluctuations inherent in a measured system, and the other is the noise caused by the process of measurement.
Quantum fluctuations prevent us from knowing a quantum system beyond the probability distribution of the measured observable~\cite{Bell}, but the probability distribution itself can be  accurately determined by means of an appropriate projection measurement. 
The uncertainty relation between two noncommuting observables such as the position and the momentum originates from those fluctuations~\cite{Kennard,Robertson,Deutsch,Maassen-Uffink}.
On the other hand, the noise places a limit on the accuracy of simultaneous measurement. 
It is known, for example, that simultaneous measurements of two noncommuting observables implies that at least one of them cannot be measured without incurring a measurement error~\cite{Neumann}. 
Despite a long history of study~\cite{Neumann,Heisenberg,Jammer,Bohr}, however, the fundamental limit to simultaneous measurement of two noncommuting observables has yet to be fully understood. 

A classic analysis of the problem was given by Arthurs~\textit{et al.}~\cite{Arthurs-Kelly,Arthurs-Goodman} who, under a special condition called unbiasedness, have shown that the lower bound of the uncertainty product of canonically conjugate observables is twice as large as the standard lower bound of $\hbar/2$, where $\hbar$ is the Planck constant divided by $2 \pi$. 
The underlying physics behind this doubling of the lower bound is that, under the condition of unbiasedness, fluctuations of a system's observable and the noise generated in the measurement process  become uncorrelated and that they simply add up.
More recently, a number of studies on related problems are conducted without invoking the unbiasedness condition and various uncertainty relations are derived~\cite{Arthurs-Kelly,Yuen-Lax,Busch,Yamamoto-Haus,Arthurs-Goodman,Appleby,Busch-Shilladay,Ozawa-1,Andersson-Barnett-Aspect,Massar,Martens-Muynck,Muynck,Hall,Werner}.

In this paper, we derive a trade-off relation concerning the measurement accuracy of two noncommuting observables for a qubit system by considering nonideal joint  measurements.  We also show that our characterization of the measurement accuracy is closely related to the Fisher information~\cite{Fisher,Cover-Thomas}, which provides the asymptotic accuracy of the maximum-likelihood estimation of the probability distribution of an observable for a finite number of samples.  In reality, only a finite number of samples are available~\cite{Massar-Popescu}, which give us only an imperfect information about the probability distribution of an observable for an unknown state.  The crucial observation made in this paper is that this imperfection is further deteriorated in the case of simultaneous measurement due to noncommutability of the observables.

We first formulate a simultaneous measurement on a qubit system with respect to two observables
$\hat A=\bm{n}_A \cdot \hat{\bm{\sigma}}$ and $\hat B=\bm{n}_B \cdot \hat{\bm{\sigma}}$, where three-dimensional unit vectors $\bm{n}_A$ and $\bm{n}_B$ indicate the directions of the measurements with $\bm{n}_A \cdot \bm{n}_B=\cos \theta \ (0< \theta < \pi)$, and  $\hat{\bm{\sigma}}\equiv(\hat \sigma_x, \hat \sigma_y , \hat \sigma_z)$ represents  of the Pauli matrices.
Since both $\hat A$ and $\hat B$ have eigenvalues of $\pm 1$, they can be represented as $\hat A = \hat{P}_A(+) - \hat{P}_A(-)$ and $\hat B = \hat{P}_B(+) - \hat{P}_B(-)$, where $\hat P_{A(B)}(\pm )$ are projection operators corresponding to observable $\hat A$ ($\hat B$).
We denote as $p_{A(B)}(i)$ the probability distribution of $\hat A (\hat B)$ given by
\begin{equation}
p_{\alpha}(i) = {\rm tr} (\hat{\rho} \hat{P}_{\alpha}(i)) \ (\alpha =A,B),
\label{3}
\end{equation}
where $\hat \rho$ is the density operator of the measured qubit system.
Each simultaneous measurement on the qubit system should yield a pair of outcomes $(i, j)$ ($i,j=\pm$) for observables $\hat A$ and $\hat B$, because their eigenvalues are $\pm 1$.
We characterize the probability distributions of obtaining $(i ,j)$ with a positive operator-valued measure (POVM) $\{ \hat E(i, j) \}$~\cite{Davis-Lewis,Nielsen-Chuang}, where $\sum_{i,j} \hat E(i, j)=\hat I$, with $\hat I$ being the identity operator.
The probability distribution $q(i,j)$ of obtaining an outcome $(i,j)$ is given by
\begin{equation}
q(i,j) = {\rm tr} (\hat{\rho} \hat{E}(i,j)),
\label{4}
\end{equation}
and the marginal POVMs are expressed by 
\begin{equation}
\hat E_A(i) = \sum_{j=+,-} \hat E(i,j), \ \hat E_B(j) = \sum_{i=+,-} \hat E(i,j).
\label{5}
\end{equation}
The four positive operators~$\hat E(i,j)$ are, in general, expressed as
\begin{equation}
\hat E(i,j)= r_{ij}\hat I+ \bm{x}_{ij}\cdot \hat{\bm{\sigma}}.
\label{19}
\end{equation}
The requirements that the sum of the four operators equals the identity operator and that all of them be non-negative are met if and only if
\begin{equation}
\sum_{i,j} r_{ij}=1, \ \sum_{i,j} \bm{x}_{ij}=\bm{0}, \ |\bm{x}_{ij}|\leq r_{ij}.
\label{21}
\end{equation}
The marginal POVMs are also expressed as
\begin{equation}
\hat{E}_{\alpha}(+) = r_{\alpha}\hat{I} + \bm{x}_{\alpha} \cdot \hat{\bm{\sigma}}, \ \hat{E}_{\alpha}(-) = \hat{I} - \hat{E}_{\alpha}(+),
\end{equation}
where  $r_A \equiv r_{++} + r_{+-}$, $r_B \equiv r_{++} + r_{-+}$, $\bm{x}_A= \bm{x}_{++}+\bm{x}_{+-}$, and $\bm{x}_B= \bm{x}_{++}+\bm{x}_{-+}$.
The probability distributions of measurement outcomes are given by
\begin{equation}
q_{\alpha}(i) = {\rm tr} (\hat{\rho} \hat{E}_{\alpha}(i)). 
\label{q}
\end{equation}
In general, $q_A(i)$ and $q_B(i)$ do not respectively coincide with $p_A(i)$ and $p_B(i)$,  because the simultaneous measurement entails a measurement error. 

In the following, we consider a class of simultaneous measurements called nonideal joint measurement~\cite{Martens-Muynck,Muynck}.  A simultaneous measurement $\hat A$ and $\hat B$ belonging to this class satisfies the condition that 
\begin{equation}
\hat{E}_{\alpha}(i) = \sum_{j=+,-} (F_{\alpha})_{ij}\hat{P}_{\alpha}(j),
\label{7}
\end{equation}
where $F_A$ and $F_B$ are two-dimensional square matrices:
\begin{eqnarray}
F_{\alpha} =
\left( 
\begin{array}{cc}
r_{\alpha} \pm |\bm{x}_{\alpha}| & r_{\alpha} \mp |\bm{x}_{\alpha}| \\
1-r_{\alpha} \mp |\bm{x}_{\alpha}| & 1-r_{\alpha} \pm |\bm{x}_{\alpha}| \\
\end{array} 
\right).
\end{eqnarray}
It can be shown that  $\bm{x}_A /\hspace{-1mm}/ \bm{n}_A$ and  $\bm{x}_B /\hspace{-1mm}/ \bm{n}_B$. 
Moreover, we can show that $F_A$ and $F_B$ are transition-probability matrices, or stochastic matrices, satisfying
\begin{equation}
0 \leq |{\rm det}F_{\alpha}| \leq 1, \ |{\rm det}F_{\alpha}| = 2|\bm{x}_{\alpha}|.
\end{equation}
Note that the nonideality of  measurement is characterized only by the transition-probability matrix in the case of  a nonideal measurement. 
A nonideal measurement expressed by Eq.~\eqref{7} is formally interpreted as a measurement of $\hat A (\hat B)$ to which classical noise, characterized by a transition-probability matrix or a noisy transmission channel, is added.  However, in general, the noise in Eq.~\eqref{7} arises from interactions described by quantum mechanics.

Let us now quantify the accuracy of  measurement.
Our purpose is to characterize the accuracy in such a manner that it depends only on the process of measurement and  not on measured state $\hat{\rho}$. To do this, we focus on the transition-probability matrix $F_{\alpha}$, which characterize the noise caused by the measurement process, and define 
\begin{equation}
\mathcal{X}_A \equiv ({\rm det}F_A)^2, \ \mathcal{X}_B \equiv ({\rm det}F_B)^2,
\label{10}
\end{equation}
where $0 \leq \mathcal{X}_A \leq 1$ and $0 \leq \mathcal{X}_B \leq 1$. We shall refer to $\mathcal{X}_{A(B)}$ as the accuracy of  measurement for $\hat A$ ($\hat B$). 

When $\mathcal{X}_A$ or $\mathcal{X}_B$ equals $1$,  the measurement of $\hat A$ or that of  $\hat B$ can be shown to be a projection measurement.
On the other hand, when $\mathcal{X}_A$ or $\mathcal{X}_B$ equals $0$, all of the operators in  $\{ \hat E_A (i) \}$ or $\{ \hat E_B (i) \}$ are proportional to the identity operator and therefore   no information about $\hat A$ or $\hat B$ can be obtained from the measurement.  Note that achieving both $\mathcal{X}_A =1$ and $\mathcal{X}_B =1$ is impossible for simultaneous measurement, because a measurement error is unavoidable  in at least  one of the two noncommuting observables.
In fact, we can derive a stronger trade-off relation between $\mathcal{X}_A$ and $\mathcal{X}_B$ :
\begin{equation}
\mathcal{X}_A + \mathcal{X}_B - \mathcal{X}_A\mathcal{X}_B \cos^2 \theta \leq 1,
\label{11}
\end{equation}
where $\theta=\cos^{-1}( \bm{n}_A \cdot \bm{n}_B)$.
A similar inequality is obtained in Ref.~\cite{Busch,Andersson-Barnett-Aspect} for a special case of $r_A=r_B=1/2$.
The accessible regime for $\mathcal{X}_A$ and $\mathcal{X}_B$ is illustrated in FIG.1 for the case of $\theta = \pi / 6$.
For example, $\mathcal{X}_A \to 1$ can be achieved only when $\mathcal{X}_B \to 0$, indicating that when we measure one observable without any measurement error, we cannot obtain any information about the other observable.
Trade-off relation~\eqref{11} between  $\mathcal{X}_A$ and $\mathcal{X}_B$ implies an upper bound on our knowledge about noncommuting observables for a qubit system.

\begin{figure}[htbp]
 \begin{center}
  \includegraphics[width=70mm]{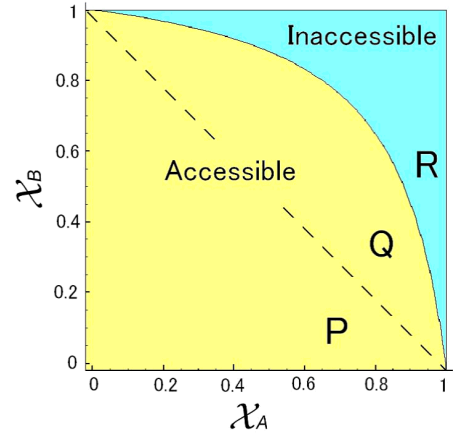}
 \end{center}
 \caption{(Color) Trade-off relation on the accuracy of noncommuting observables. P and Q indicate the regimes satisfying inequality~\eqref{11} for the case of $\theta=\pi/6$. The regime Q is accessible only through  simultaneous measurement.}
 \label{fig:Trade-off}
\end{figure}

The derivation of inequality~\eqref{11} goes as follows. From Eqs.~\eqref{21}, we have
\begin{equation}
\begin{split}
|\bm{x}_A+\bm{x}_B-\bm{y}|\leq 2r_{++}, \ |\bm{x}_A-\bm{x}_B+\bm{y}|\leq 2r_{+-}, \\
|\bm{x}_A-\bm{x}_B-\bm{y}|\leq 2r_{-+}, \ |\bm{x}_A+\bm{x}_B+\bm{y}|\leq 2r_{--},
\label{22}
\end{split}
\end{equation}
where $\bm{y} \equiv \bm{x}_{+-}+\bm{x}_{-+}$.
By using \eqref{22} and the triangle inequality, we obtain $|\bm{x}_A+\bm{x}_B|+  |\bm{x}_A-\bm{x}_B| \leq 1$.
Noting that $| {\rm det}F_A | = 2|\bm{x}_A|, \ |{\rm det}F_B| = 2|\bm{x}_B|$ and $\bm{x}_A \cdot \bm{x}_B=|\bm{x}_A| |\bm{x}_B| \cos \theta$, we obtain \eqref{11}.

We give an alternative expression of inequality~\eqref{11}~\cite{Andersson-Barnett-Aspect}.
Let us define $\mathcal{E}_A \equiv (1/\mathcal{X}_A) - 1$ and $\mathcal{E}_B \equiv (1/\mathcal{X}_B) - 1$, where $0 \leq \mathcal{E}_A < \infty$ and $\ 0 \leq \mathcal{E}_B < \infty$.
Here, $\mathcal{E}_A=0$ implies $\mathcal{X}_A=1$, that is, the measurement of observable $\hat A$ involves no measurement error.
Hence, we can interpret $\mathcal{E}_A$ and $\mathcal{E}_B$ as parameters characterizing measurement errors.
In terms of $\mathcal{E}_A$ and $\mathcal{E}_B$, we can express \eqref{11} as
\begin{equation}
\mathcal{E}_A \mathcal{E}_B \geq \sin^2 \theta.
\label{12}
\end{equation}
This inequality manifestly shows that a trade-off relation exists between measurement errors $\mathcal{E}_A$ and $\mathcal{E}_B$ for simultaneous measurement if the corresponding observables are noncommuting (i.e. $\theta \ne 0$).

An optimal simultaneous measurement that satisfies the equality in \eqref{11} with $\mathcal{X}_A=\mathcal{X}_B$  is achieved when $\bm{x}_{++}+\bm{x}_{--}=0$, $\bm{x}_{+-}+\bm{x}_{-+}=0$, $|\bm{x}_{++}|=r_{++}=r_{--}$, $|\bm{x}_{+-}|=r_{+-}=r_{-+}$, and $r_{++}+r_{+-}=1/2$. 
The corresponding POVM $\{ \hat E(i,j) \}$ is given by $\hat E(i,j)= |\bm{x}_{ij} | \hat I+ \bm{x}_{ij}\cdot \hat{\bm{\sigma}}$ with
\begin{equation}
\begin{split}
\bm{x}_{++}=r(\bm{n}_A+\bm{n}_B), \ \bm{x}_{-+}= - \bm{x}_{+-}, \\
\bm{x}_{+-}=r(\bm{n}_A-\bm{n}_B), \ \bm{x}_{++}= - \bm{x}_{--},
\end{split}
\label{13}
\end{equation}
where $r= \pm \left(|\bm{n}_A+\bm{n}_B|+ |\bm{n}_A-\bm{n}_B| \right)^{-1}/2$. 
These positive operators are proportional to projection operators, and the directions of the projection are mutually orthogonal~\cite{Cohen-Scully}. 

Inequalities~\eqref{11} and~\eqref{12} offer a rigorous representation of Bohr's principle of complementarity. According to his classic paper~\cite{Bohr}, the essence of the principle of complementarity is ``the mutual exclusion of any two experimental procedures'' when measuring two noncommuting observables simultaneously.  Inequalities~\eqref{11} and~\eqref{12} quantitatively represent this incompatibility.

We next point out a close connection between the accuracy of measurement $\mathcal{X}_{\alpha}$ and the asymptotic accuracy of the maximum-likelihood estimation of $p_{\alpha} (i)$.  Let us consider simultaneous measurements for each of $N$ ($<\infty$) samples prepared in the same unknown state $\hat{\rho}$. The issue here is how accurately we can estimate the true probability distributions of $\hat A$ and $\hat B$ from a finite number of samples.
The condition~\eqref{7} of the nonideal joint measurement is equivalent to the condition that
\begin{equation}
q_{\alpha}(i) = \sum_{j=+,-} (F_{\alpha})_{ij}p_{\alpha}(j)
\label{6}
\end{equation}
holds for any $\hat \rho$.
In this case, probability distributions $q_{\alpha}(i)$ is parameterized by true probability distributions $p_{\alpha}(i)$, so that we can estimate $p_{\alpha}(i)$ by using the method of the classical maximum-likelihood estimation  except for the case of $\bm{x}_A = \bm{0}$ or $\bm{x}_B = \bm{0}$.  We assume that $\bm{x}_A \neq \bm{0}$ and $\bm{x}_B \neq \bm{0}$ in the following analysis.

Suppose that we obtain outcome ``$+$'' $N_{\alpha}(+)$ times and outcome ``$-$'' $N_{\alpha}(-)$ times, where $N_{\alpha}(+)+N_{\alpha}(-)=N$. The likelihood function then becomes
\begin{equation}
L_{\alpha}\left( p_{\alpha}(+) \right) = \sum_{i=+,-} N_{\alpha}(i) \ln q_{\alpha}(i).
\end{equation}
We denote  as $p_{\alpha}^{\ast}$ the maximum-likelihood estimator of $p_{\alpha}(+)$ inferred from $N$ measurement outcomes such that $L_{\alpha}\left( p_{\alpha}(+) \right)$ takes the maximum value with $p_{\alpha}(+) = p_{\alpha}^{\ast}$ under  condition $0 \leq p_{\alpha} (+) \leq 1$.
The accuracy of the estimate can then be asymptotically characterized by the following theorem~\cite{Fisher,Cover-Thomas}:
\textit{the distribution of $p_{\alpha}^{\ast}$ approaches the normal distribution with average $p_{\alpha}(+)$ and variance $(NI_{\alpha})^{-1}$ as the number of samples $N$ increases with $I_{\alpha}$ being the Fisher information.}
Note that this theorem does not hold if $p_{\alpha}(+)=0$ or $1$. 
In other words, the  maximum-likelihood estimator $p_{\alpha}^{\ast}$ asymptotically approaches true value $p_{\alpha}(+)$,  the asymptotic behavior being characterized by Fisher information $I_{\alpha}$.
The larger the Fisher information, the more information we can extract from the measurement outcome.
In the special case of $I_{\alpha}=0$, no information about the measured system can be obtained from measurement.
Once we have found $p_{\alpha}(+)$,  $p_{\alpha}(-)$ is obtained from $p_{\alpha}(-)=1-p_{\alpha}(+)$.
In the present situation, the Fisher information is given by
\begin{equation}
\begin{split}
I_{\alpha} &\equiv  - \sum_{i=+,-} q_{\alpha} (i) \frac{\partial^2 \ln q_{\alpha}(i)}{\partial p_{\alpha}(+)^2} \\
&= \frac{\mathcal{X}_\alpha}{q_{\alpha}(+)q_{\alpha}(-)}.
\end{split}
\label{9}
\end{equation}
Note that in simultaneous measurement the Fisher information on observable $\hat{A}$ ($\hat B$) is deteriorated by a factor of $\mathcal{X}_{A(B)}$ in comparison with projection measurement.

We thus conclude that the trade-off relation~\eqref{11} or \eqref{12} indicates that knowing about one of the probability distributions of the two noncommuting observables prevents us from knowing about the other because the number of available samples is finite; the upper bound on the estimation accuracy of the probability distributions is severely restricted, unless the number of available samples is infinite.  Note that if $N$ is infinite, we can accurately reconstruct both $p_{A}(i)$ and $p_{B}(i)$ from measured probability distributions $q_{A}(i)$ and $q_B(i)$; with infinite samples, the noncommutability does not affect the accuracy of the estimates of probability distributions of noncommuting observables.

We have considered the estimates of probability distributions $p_A(i)$ and $p_B(i)$ for simultaneous measurement by using the POVM which is made up by  four positive operators.
We now discuss a simple strategy for estimating the probability distributions of noncommuting observables.
We divide $N$ identically prepared samples into two groups according to the ratio $\xi : 1-\xi \ (0<\xi <1)$, and estimate $p_A(i)$ by  measuring  $\hat A$ for the first group using POVM $\{ \hat E_A (\pm) \}$, and similarly we estimate $p_B(i)$ for the second group using POVM $\{ \hat E_B (\pm) \}$.
For a large number of samples, this measurement is asymptotically  described by POVM $ \{ \xi \hat E_A(\pm), (1-\xi )\hat E_B(\pm ) \}$. In this case, the accuracy of the measurement is limited by $\mathcal{X}_A \leq \xi$ and $\mathcal{X}_B \leq 1-\xi$, because  the accuracy per sample is deteriorated by a factor of $\xi$ and $1-\xi$, respectively. We thus  obtain $\mathcal{X}_A + \mathcal{X}_B \leq 1$ (domain P in FIG.1), and therefore conclude that simultaneous measurement has an advantage over this simple method in that the domain $\mathcal{X}_A + \mathcal{X}_B > 1$ for $\theta \ne \pi/2$, i.e. domain Q in FIG.1 is accessible.

It is worth pointing out that we can interpret the trade-off relations~\eqref{11} and \eqref{12} as uncertainty relations between the measurement error and the back-action of the measurement~\cite{Heisenberg,Appleby,Banaszek-Devetak,Ozawa-2,Fuchs,Busch-Heinomen-Lahti,Werner}.
Let us suppose that $\hat{\rho}'$ is a state immediately after measurement of $\hat A$ for  premeasurement state $\hat \rho$. 
To identify the disturbance of $\hat B$ caused by $\hat A$, we consider how much information about $\hat B$ for premeasurement state $\rho$ is left in postmeasurement state $\hat{\rho}'$.
We characterize this by considering how much information on $\hat{\rho}$  can be obtained by measuring $\hat B$ for $\hat{\rho}'$.
The joint operation of measurement $\hat A$ being followed by measurement $\hat B$ can be described by a  POVM $\{ \hat{E}(i,j) \}$ corresponding to the measurement outcomes $(i,j)$~\cite{Nielsen-Chuang}. 
If the POVM fulfills the requirement that its marginal POVMs describe nonideal measurements of $\hat A$ and $\hat B$, we again obtain inequality~\eqref{11}.
It is possible to interpret $1- \mathcal{X}_B$ as a measure of the back-action of $\hat B$ caused by measurement of $\hat A$.
Defining the measurement error of $\hat A$ as $\mathcal{E}_A \equiv (1/\mathcal{X}_A) - 1$ and the back-action of the measurement on $\hat B$ as $\mathcal{D}_B \equiv (1/\mathcal{X}_B) - 1$, we obtain
\begin{equation}
\mathcal{E}_A \mathcal{D}_B \geq \sin^2 \theta.
\label{14}
\end{equation}
We should note that a non-selective measurement process for $\hat A$ can simulate the decoherence caused by the environment. 
In this case, trade-off relations~\eqref{11} and \eqref{14} give a lower bound to the back-action of $\hat B$ in the presence of  decoherence characterized by $\mathcal{X}_A$.

Note that if it were possible to perform error-free measurement of $\hat A$ without disturbing $\hat B$, then we could precisely measure both $\hat A$ and $\hat B$ by performing measurement $\hat A$  followed by measurement $\hat B$. 
We can say that  the formal similarity of two kinds of trade-off relations,  \eqref{11} (or equivalently \eqref{12}) and \eqref{14}, quantitatively represents the above-mentioned connection between the two kinds of uncertainty relations~\cite{Appleby,Ozawa-1}: the uncertainty relation between the error of $\hat A$ and that of $\hat B$, and the uncertainty relation between the error of $\hat A$ and the back-action of $\hat B$. 
In the context of the maximum-likelihood estimation with finite samples, the number of samples required to achieve a given accuracy in estimating the probability distributions of both $\hat A$ and $\hat B$ is greater than the number for the case of estimating either $\hat A$ or $\hat B$,  because samples on which we perform precise measurements of $\hat A$ lose the information about the probability distribution of $\hat B$, due to the back-action of measurement $\hat A$.

Finally, we mention the relevance of our work to previous work in the case of a qubit system.
Arthurs and Goodman~\cite{Arthurs-Goodman} have discussed simultaneous measurement by assuming the unbiasedness for observables $\hat A$ and $\hat B$.
This condition implies that the arithmetic average of the measurement outcomes approaches to the true average in the limit of $N \to \infty$.
In the case of a qubit system, the unbiasedness condition implies $\mathcal{X}_A = \mathcal{X}_B=1$ because the probability distribution has a one-to-one correspondence with the average value.
As discussed earlier, however, no simultaneous measurement satisfies this condition.
We therefore cannot address simultaneous measurement by the method of Arthurs and Goodman for a qubit system. 
On the other hand, Andersson \textit{et al.}~\cite{Andersson-Barnett-Aspect} have relaxed the unbiasedness condition and adopted the condition that the arithmetic average of the measurement outcomes is proportional to the true average in the limit of $N \to \infty$.
This condition is satisfied if and only if $r_A = r_B = 1/2$.
In this paper, we have considered the most general class of measurements to which we can use the maximum-likelihood estimation to the marginal probability distributions, without assuming the unbiasedness condition.  This class of measurements has been studied by de Muynck and Martens who have derived a trade-off relation in terms of the Shannon channel capacity~\cite{Martens-Muynck,Muynck}.    

In conclusion, we have derived a trade-off relation concerning the accuracy of simultaneous measurement of two noncommuting observables for a qubit system. The relation gives an upper bound on our knowledge about noncommuting observables.
Moreover, we have pointed out that the accuracy parameter $\mathcal{X}_{A(B)}$ is quantitatively related to the asymptotic accuracy of the maximum-likelihood estimation of the probability distribution of observable $\hat A (\hat B)$.
The generalization of our results to high-spin systems and continuous-variable systems merits further study.

This work was supported by a Grant-in-Aid for Scientific Research (Grant No. 17071005) and by a 21st Century COE program at Tokyo Tech, ``Nanometer-Scale Quantum Physics'', from the Ministry of Education, Culture, Sports, Science and Technology of Japan. M.U. acknowledges support by a CREST program of JST.


\begin{thebibliography}{99}
\bibitem{Bell}
J. S. Bell, Physics \textbf{1}, 195 (1964).

\bibitem{Kennard}
E. H. Kennard, Z. Phys. \textbf{44}, 326 (1927).

\bibitem{Robertson}
H. P. Robertson, Phys. Rev. \textbf{34}, 163 (1929).

\bibitem{Deutsch}
D. Deutsch, Phys. Rev. Lett. \textbf{50}, 631 (1983). 

\bibitem{Maassen-Uffink}
H. Maassen and J. B. M. Uffink, Phys. Rev. Lett. \textbf{60}, 1103 (1988).

\bibitem{Neumann}
J. von Neumann, \textit{Mathematical Foundations of Quantum Mechanics} (Princeton Univ. Press, Princeton, 1955). 

\bibitem{Heisenberg}
W. Heisenberg, Z. Phys. \textbf{44}, 172 (1927).

\bibitem{Bohr}
N. Bohr, Phys. Rev. \textbf{48}, 696 (1935).

\bibitem{Jammer}
M. Jammer, \textit{The philosophy of quantum mechanics} (Wiley, New York, 1974).

\bibitem{Arthurs-Kelly}
E. Arthurs and J. L. Kelly, Jr., Bell. Syst. Tech. J. \textbf{44}, 725 (1965).


\bibitem{Arthurs-Goodman}
E. Arthurs and M. S. Goodman, Phys. Rev. Lett. \textbf{60}, 2447 (1988).

\bibitem{Yuen-Lax}
H. P. Yuen and M. Lax, IEEE Trans. Inf. Theory \textbf{IT-19}, 740 (1973).

\bibitem{Busch}
P. Busch, Phys. Rev. D \textbf{33}, 2253 (1986).

\bibitem{Yamamoto-Haus}
Y. Yamamoto and H. A. Haus, Rev. Mod. Phys. \textbf{58}, 1001 (1986).


\bibitem{Martens-Muynck}
H. Martens and W. M. de Muynck, Found. Phys. \textbf{20}, 255 (1990). 

\bibitem{Appleby}
D. M. Appleby, Int. J. Theor. Phys. \textbf{37} 2557 (1998).

\bibitem{Muynck}
W. M. de Muynck, Found. Phys. \textbf{30}, 205 (2000).

\bibitem{Busch-Shilladay}
P. Busch and C. R. Shilladay, Phys. Rev. A \textbf{68}, 034102 (2003).

\bibitem{Hall}
M. J. W. Hall, Phys. Rev. A, \textbf{69}, 052113 (2004).

\bibitem{Ozawa-1}
M. Ozawa, Phys. Lett. A \textbf{320}, 367 (2004). 

\bibitem{Werner}
R. Werner, Qu. Inf. Comp. \textbf{4} 546 (2004).

\bibitem{Andersson-Barnett-Aspect}
E. Andersson, S. M. Barnett and A. Aspect, Phys. Rev. A \textbf{72}, 042104 (2005).

\bibitem{Massar}
S. Massar, e-Print: quant-ph/0703036 (2007).


\bibitem{Fisher}
R. A. Fisher, Proc. Camb. Phil. Soc. \textbf{22}, 700 (1925).

\bibitem{Cover-Thomas}
T. M. Cover and J. A. Thomas, \textit{Elements of Information theory} (John Wiley and Sons, 1991).

\bibitem{Massar-Popescu}
S. Massar and S. Popescu, Phys. Rev. Lett. \textbf{74} 1259 (1995).


\bibitem{Davis-Lewis}
E. B. Davies and J. T. Lewis, Commun. Math. Phys. \textbf{17}, 239 (1970).

\bibitem{Nielsen-Chuang}
M. A. Nielsen and I. L. Chuang, \textit{Quantum Computation and Quantum Information} (Cambridge University Press, Cambridge, 2000).

\bibitem{Cohen-Scully}
L. Cohen and M. O. Scully, Found. Phys. \textbf{16}, 295 (1986).

\bibitem{Fuchs}
C. A. Fuchs, Fortschr. Phys. \textbf{46},  535 (1998).

\bibitem{Banaszek-Devetak}
K. Banaszek and I. Devetak, Phys. Rev. A \textbf{64}, 052307 (2001).

\bibitem{Ozawa-2}
M. Ozawa, Ann. Phys. \textbf{311}, 350 (2004).

\bibitem{Busch-Heinomen-Lahti}
P. Busch, T. Heinomen and P. J. Lahti, Phys. Lett. A \textbf{320}, 261 (2004).

\end{thebibliography}
\end{document}